# A Multi-level Clustering Approach for Anonymizing Large-Scale Physical Activity Data

Pooja Parameshwarappa, *University of Maryland, Baltimore County, USA*
Zhiyuan Chen, *University of Maryland, Baltimore County, USA*
Güneş Koru, *University of Maryland, Baltimore County, USA*

**ABSTRACT**

*Publishing physical activity data can facilitate reproducible health-care research in several areas such as population health management, behavioral health research, and management of chronic health problems. However, publishing such data also brings high privacy risks related to re-identification which makes anonymization necessary. One of the challenges in anonymizing physical activity data collected periodically is its sequential nature. The existing anonymization techniques work sufficiently for cross-sectional data but have high computational costs when applied directly to sequential data. This paper presents an effective anonymization approach, Multi-level Clustering based anonymization to anonymize physical activity data. Compared with the conventional methods, the proposed approach improves time complexity by reducing the clustering time drastically. While doing so, it preserves the utility as much as the conventional approaches.*

## INTRODUCTION

There has been a rapid increase in the availability of physical activity data due to the increase in the use of wearable devices, smartphones, and smart environments. Publishing physical activity data can support reproducible research in personal and population health management, behavioral health research and management of chronic health problems. For example, data about vigorous activity and sedentary hours per day can help research studies investigating the types and amounts of physical activity necessary at the individual, cohort and population levels (Matthews et al., 2008; Pate et al., 1995). Physical activity is known to decrease the risk of various diseases such as cardiovascular diseases, diabetes and obesity (Dietz, Douglas, & Brownson, 2016; Thornton et al., 2016). Publishing activity data can support research in preventing such chronic diseases. Furthermore, it can facilitate research studies that aim to reduce health care costs and the costs related to social benefits and work absenteeism (CDC Foundation, 2015; Spenkelink, Hutten, Hermens, & Greitemann, 2002). Therefore, there is an important and increasing need for publishing physical activity data.

However, publishing physical activity data also brings high privacy risks related to re-identification. Although direct identifiers such as names, identification numbers, and other personally identifiable information (PII) are removed, many unique longitudinal patterns can easily reveal identities. For example, consider the publication of a data set which includes activity data of a group of people and their health status. Table 1 shows an example which contains activity data for four individuals collected every minute for a certain time duration. Additionally, the data contains health status of these individuals. Assume that, an adversary gets access to this data and knows that an individual whose record is in the data runs every Monday, Tuesday, and Wednesday at 6:00 am. Since there is only one person with this specific routine, his/her data is easily re-identifiable. As a result, the adversary gains access to sensitive information such as the health status. To reduce the probability of re-identification to acceptable levels, and ensure privacy, such activity data needs to be anonymized. Anonymization involves modifying the data, in order to protect the privacy of the individuals whose information is in the data, while preserving the utility of the data.

*Table 1. Example Showing Physical Activity Data of Four People and Corresponding Health Status. S stands for Stationary, W stands for Walking and R stands for Running*

|  | Physical Activity Data | | | | | | | | Health Status |
|---|---|---|---|---|---|---|---|---|---|
| Day | Mon | Mon | .. | Tue | Tue | .. | Wed | Wed | .. | |
| Time | 6:00 am | 6:01 am | .. | 6:00 am | 6:01 am | .. | 6:00 am | 6:01 am | .. | |
| Person 1 | S | S | .. | S | W | .. | S | S | .. | Heart Disease |
| Person 2 | R | R | .. | R | R | .. | R | R | .. | Depression |
| Person 3 | S | S | .. | S | S | .. | S | S | .. | Cold |
| Person 4 | S | S | .. | S | S | .. | W | W | .. | Heart Disease |

*Table 2. Sample Cross-Sectional Data*

|  | Age | Sex | Zipcode | Disease |
|---|---|---|---|---|
| Person 1 | 22 | M | 21220 | Cold |
| Person 2 | 25 | F | 21222 | Heart Disease |
| Person 3 | 33 | F | 21236 | Cancer |
| Person 4 | 30 | M | 21239 | Cancer |

Unfortunately, most of the conventional anonymization techniques are suitable for cross-sectional data sets (El Emam et al., 2009; Gal, Chen, & Gangopadhyay, 2008; Loukides, Gkoulalas-Divanis, & Malin, 2010). An example for cross-sectional data is shown in Table 2. However, physical activity data is sequential in nature (shown in Table 1), which results in high dimensionality because every instance of time acts as a dimension. For example, if the data is collected every minute for a period of one month, then the number of dimensions would be 43,200.

Furthermore, most of the existing techniques for anonymizing sequential data sets (Gkoulalas-Divanis & Loukides, 2012; He, Cormode, Machanavajjhala, Procopiuc, & Srivastava, 2015; Martinez-Bea & Torra, 2011; Pensa, Monreale, Pinelli, & Pedreschi, 2008) focus on preserving frequent sequential patterns to maintain the utility of the data. On the other hand, for physical activity data, preserving aggregate statistics is more relevant than sequential pattern-preservation for meeting the utility requirements (Matthews et al., 2008; Spees, Scott, & Taylor, 2012). Additionally, most of the existing techniques do not consider the computational efficiency of the anonymization method, which is important for high dimensional data sets. Applying the existing techniques directly to high dimensional data results in very high computational costs.

In this paper, the authors propose a Multi-level Clustering (MC) based anonymization approach that improves time complexity by reducing the clustering time drastically. The approach focuses on preserving the aggregate statistics and correlations in the data in order to maintain the utility of the data. The approach includes two steps:

1. Clustering physical activity data using MC so that activity sequences that are similar to one another belong to the same cluster

2. Application of k-anonymity (Sweeney, 2002b) and differential privacy (Dwork, 2008; Dwork, McSherry, Nissim, & Smith, 2006) models on the clusters generated, thereby resulting in Multi-level Clustering based K-Anonymity (MCKA) and Multi-level Clustering based Differential Privacy (MCDP), respectively.

The rest of the paper is organized as follows. The next section presents the related work. It is followed by the methods section describing the proposed methodology. The next section describes the experiments and results, followed by the discussion section and the conclusion section.

# RELATED WORK

Some of the commonly used privacy models include *k*-anonymity (Sweeney, 2002b), *l*-diversity (Machanavajjhala, Kifer, Gehrke, & Venkitasubramaniam, 2007), *t*-closeness (Li, Li & Venkatasubramanian, 2007) and differential privacy (Dwork, 2008; Dwork et al., 2006). Some of the techniques for achieving these privacy models include generalization and suppression (Sweeney, 2002a), perturbative techniques such as microaggregation (Domingo-Ferrer, Martínez-Ballesté, Mateo-Sanz, & Sebé, 2006a; Domingo-Ferrer & Torra, 2005; Templ, Meindl, & Kowarik, 2013) and Laplace Perturbation Algorithm (Dwork et al., 2006). The following sub-sections present the existing work for anonymizing cross-sectional and sequential data. At the end of the section, research gap addressed by the proposed approach is presented.

## Anonymization of Cross-sectional Data

Most of the conventional anonymization techniques are suitable for cross-sectional data sets. One of the approaches for anonymizing cross-sectional health data is Optimal Lattice Anonymization (OLA) (El Emam et al., 2009), which provides a globally optimal anonymization solution through generalization. El Emam (2008), and El Emam and Dankar (2008) present various heuristics for de-identifying cross-sectional health data, re-identification risks associated with using k-anonymization for health data, and improvements to control the risk. Another approach anonymizes data containing both clinical and genomic information by extracting linkable features from clinical data and generalizing these features such that it is no longer possible to link the genomic data with a small number of patients (Loukides et al., 2010). Another anonymization approach extends k-anonymity and l-diversity models for anonymizing patient data with multiple sensitive attributes (Gal et al., 2008). Gal et al. (2014) propose an anonymization framework which identifies a suitable de-identification method for cross-sectional data based on the requirements of the recipient of the data. Poulis et al. (2017) propose anonymization of health data that contains both demographics and diagnoses codes. The authors use (k, $k^m$) anonymity which assumes that the attacker knows the demographics and up to *m* diagnoses codes of a patient. This approach takes into consideration the utility constraint set and provides low information loss. Zhang et al. (2014) propose anonymization of big data using MapReduce (Dean & Ghemawat, 2010). This approach combines Top–Down Specialization (TDS) and Bottom–Up Generalization (BUG) components for anonymization. It is a hybrid approach that automatically chooses the component suitable for anonymization and is implemented using the MapReduce paradigm. A two-phase clustering method using Map-Reduce was proposed in (Zhang, Dou, Pei, Nepal, Yang, Liu & Chen, 2014) where a k-means like clustering is used in the first phase to generate some initial clusters and then each cluster is further divided into smaller clusters using agglomerative clustering.

Applying conventional techniques directly to sequential data is computationally expensive because of the high-dimensional nature of the sequential data. Although the approaches for big data using Map-Reduce address the issue of large-scale data, they focus only on cross-sectional data sets. In the following sub-sections, the existing techniques for anonymizing sequential data are discussed.

## Anonymization of Trajectory Data

One of the techniques for anonymizing location information data uses microaggregation as the protection method (Martinez-Bea & Torra, 2011; Nin & Torra, 2009). In microaggregation, euclidean distance and short time series distance (Martinez-Bea & Torra, 2011) are used as distance measures and information loss is determined by using measures such as average, autocorrelation (Dunn & Davis, 2017), and the difference between the original time series and the protected time series. Nergiz et al. (2008) and Nergiz et al. (2007) proposed a technique for anonymizing trajectory data based on generalization. It has two components: 1) Ensuring k-anonymity i.e. every trajectory is made indistinguishable from k-1 other trajectories; 2) Reconstruction, in which atomic trajectories are randomly sampled from the area covered by the anonymized trajectories. Mohammed et al. (2009) proposed the LKC privacy model for anonymizing trajectory data. In this model, it is assumed that the adversary knows a sub-sequence of location and timestamp pairs that the victim visited (L) and a trajectory database is said to satisfy LKC privacy if and only if for any sub-sequence, k-anonymity (K) is preserved and the probability of obtaining a sensitive value associated with this sub-sequence is below a custom threshold (C). This is achieved through coarsening in which one or more trajectory points are removed. Differentially Private Trajectory Synthesis (DPT) (He et al., 2015) is another privacy protection model in which trajectory data is represented using reference systems of different resolutions. A prefix tree is constructed for each reference system. Laplace noise is added to a subset of prefix trees, which are then used to simulate trajectories that are differentially private. Dong and Pi (2018) propose a privacy-preserving algorithm for trajectories based on frequent path. Infrequent roads are removed, and trajectories are grouped based on similarity and k-anonymity is applied to these groups. Gao et al. (2014) propose a personalized anonymization model that takes trajectory angle into consideration while creating trajectory k-anonymity set. Hu et al. (2018) propose an anonymization technique for trajectory data in which equivalence classes of trajectories are created considering the fact that different users may have different privacy requirements at different time. Barack et al. (2016) propose a semantic cloaking-based anonymization framework in which exact locations are replaced by semantic categories such as home, work, and so on.

## Anonymization of Transaction Data Sets

A variation of k-anonymity, $k^m$-anonymity (Terrovitis, Mamoulis, & Kalnis, 2008) is a privacy model for anonymizing transaction data. It is assumed that the maximum knowledge of an adversary is at most *m* items, and it is ensured that for any possible set of *m* items or less, there are at least *k* transactions. Gkoulalas-Divanis and Loukides (2012) propose two other algorithms for anonymizing transaction data sets: Privacy-Constrained Clustering-based Transaction Anonymization (PCTA) and Utility-guided Privacy-constrained Clustering-based Transaction Anonymization (UPCTA). These methods are based on agglomerative clustering. The clusters are merged at different levels, based on certain privacy (PCTA) and utility (UPCTA) constraints. Wang and Li (2018) propose anonymization of transaction data that contain sensitive information in both relational and transactional attributes. They use graph-based approach for anonymization. In this approach, the associations between customers and products are represented using an uncertain graph. The method protects the multifold privacy of the data while maximizing its utility. In another work, Kakatkar and Spann (2019) highlight the importance of anonymized and fragmented data in retailing and present a methodology to analyze such data.

## Anonymization of Other Sequential Data Sets

Randomization is one of the techniques for anonymizing time-series data (Moon, Kim, Kim, & Bertino, 2010). The original time-series is distorted by adding noise such that, the distance orders are preserved even after distortion. (k, P)-anonymity model (Shou, Shang, Chen, Chen, & Zhang, 2013) is another technique for anonymizing time-series data which ensures anonymity on two levels: k-anonymity is ensured for the entire data set, and P-anonymity is required for the pattern representations associated with the data points in the same group.

There has been some work on addressing privacy issues related to sensor data (Cavoukian, Mihailidis, & Boger, 2010; Chan & Perrig, 2003; He, Liu, Nguyen, Nahrstedt, & Abdelzaher, 2007; Li, Lou, & Ren, 2010; Sun, Fang, & Zhu, 2010). Most of these studies focus on security, access control, and encryption. A framework for anonymizing longitudinal Electronic Medical Record (EMR) data (Tamersoy, Loukides, Nergiz, Saygin, & Malin, 2012) ensures k-anonymity through generalization and suppression of International Classification of Diseases (ICD) codes and age details of patients. Longitudinal OLA (LOLA) (El Emam et al., 2012) is an extension of OLA which is used to anonymize longitudinal claims data. A system for anonymized public health data collection and intervention is proposed by Clarke and Steele (2014). In this system, statistics about the physical activity data and nutritional data from users' smartphones are computed at different resolutions based on disclosure risks and submitted to the central server. Another area of research uses Natural Language Processing (NLP) based techniques for de-identifying longitudinal clinical narratives (Dernoncourt, Lee, Uzuner, & Szolovits, 2017; Stubbs, Filannino, & Uzuner, 2017; Stubbs & Uzuner, 2015).

## Research Gap

Table 3 summarizes the advantages and limitations of the existing methods. So far, no existing technique has dealt with anonymizing physical activity data. The existing techniques discussed may not be suitable when applied directly for anonymizing such data. This is because most of these techniques focus on retaining the sub-sequences for preserving the utility of the data. However, for physical activity data, aggregate statistics such as duration, frequency, and intensity, are more relevant from the utility perspective. Also, most of the existing techniques do not consider the efficiency of the anonymization method, which is important for high dimensional data sets. To address these gaps, the authors propose an efficient approach for anonymizing physical activity data using MC and ensuring $k$-anonymity and $\varepsilon$-differential privacy, while preserving the utility of the data.

*Table 3. Advantages and Limitations of the Existing Anonymization Approaches*

| Method | Advantages | Limitations |
| --- | --- | --- |
| Anonymization of cross-sectional data | Suitable for cross-sectional data | Computationally expensive when applied to high dimensional sequential data |
| Anonymization of trajectory and transaction data | Suitable for sequential data | Focus on preserving frequent patterns. (For physical activity data, preserving aggregate statistics is more relevant) |
| Anonymization of other sequential data | Suitable for sequential data | Do not focus on the computational efficiency for high dimensional data |

## MATERIALS AND METHODS

This section first presents the dataset used for the experiments. It is followed by the explanation of the proposed approach.

### Data

Most publicly available activity datasets are small for various reasons such as unavailability of participants and need for data collection over long periods of time (Mendez-Vazquez, Helal, & Cook, 2009). However, evaluation of privacy-preserving approaches typically requires large data sets to achieve desired levels of accuracy (Kitamura, Chen, & Pendyala, 1997; Monekosso & Remagnino, 2009). To address this problem, a larger synthetic data set was generated from *Student Life dataset* (Wang et al., 2014) by paying attention to preserving the distributional characteristics of the original data.

The original dataset (*Student Life dataset*) consists of activity sequences of 49 students collected through a sensing application over a period of 10 weeks. Data for each student is a sequence of timestamp and activity inference pairs such as <1364356858; *stationary*>. This shows that, on March 27, 2013, at 04:00:58 AM, the student was stationary. Three types of activities are included: *running, walking* and *stationary*. For the same students, academic performance data, as well as behavioral and mental well-being data such as flourishing scale (i.e., self-perceived success in terms of relationships, self-esteem, purpose, and optimism) (Diener et al., 2010) are also available. Therefore, in this study, the correlations between (a) activity and Cumulative Grade Point Average (CGPA), and (b) activity and flourishing scale are used to represent utility-preservation capabilities of the proposed approach.

The synthetic data is generated by using a markov chain model (Gagniuc, 2017) in the R statistical environment (Spedicato, 2016). In the original data, activity inference is made every 2 seconds which results in very high dimensionality (approximately 450,000). In the synthetic data generation, this type of high dimensionality results in very large un-manageable state transition matrices. Also, the analyses of activity data are typically conducted at higher time levels such as daily and weekly (Matthews et al., 2008; Pate et al., 1995; Spees et al., 2012; Wang et al., 2014). Therefore, for this study, each activity sequence is aggregated to the level of one-minute intervals, and the most dominant activity during an interval is assigned as the activity for that interval. This results in 1,440 intervals a day.

In addition to the three activities, unknown values are also present. An unknown value is represented as an activity as well, and it is referred to as *missing*. This is because an adversary can make inferences based on the unknown data. For example, if the adversary knows that the victim turns the device off during class hours, the data set will contain unknown values consistently during the class hours. If there are only few records with this routine, the identity of the student is at stake.

A state transition matrix is constructed for every hour, for each student. The state transition matrices are used to simulate activity sequences for 336 hours at minute-level granularity. A total of 9,800 students' activity sequences was generated where each sequence is of length 20,160 (336 x 60). At each hour, a different student's state transition matrix is chosen at random for simulation with a probability of 0.01.

The difference between the probability distributions of the original data and the simulated data is measured using KL-divergence (Kullback & Leibler, 1951), and its value is 0.03. The low value of KL divergence indicates that the two distributions have similar behavior.

## Proposed Approach

One of the most widely used methods for achieving k-anonymity is Microaggregation (Domingo-Ferrer et al., 2006a; Domingo-Ferrer & Torra, 2005). It has two steps:

1. Partitioning: the records are partitioned into clusters of size k.

2. Aggregation: the records in each cluster are replaced by the result of an aggregation operation leading to k-anonymity

The proposed approach is based on microaggregation and has the following two steps:

1. Multi-level Clustering (MC): sequences are partitioned into clusters of size k. Clusters are homogenous i.e. each cluster contains sequences that are similar to one another. This in turn helps preserve the utility of data.

2. Anonymization (similar to aggregation step): here two types of privacy models are applied to the clusters from step 1

    a. K-anonymity, resulting in MCKA (Multi-level Clustering based K-Anonymity)

    b. Differential Privacy, resulting in MCDP (Multi-level Clustering based Differential Privacy)

### Multi-level Clustering

One of the best-known heuristic methods for achieving the partitioning step of microaggregation is Maximum Distance to Average Vector (MDAV) (Domingo-Ferrer, Solanas, & Martinez-Balleste, 2006b; Solanas, Martinez-Balleste, & Domingo-Ferrer, 2006). MDAV involves the following steps:

1. Compute the centroid of the data set. Find a point *r* that is most distant to the centroid. Find a point *s* that is farthest to *r*

2. Find $k-1$ nearest data points around both $r$ and $s$

3. If there are at least $2k$ data points remaining, then repeat steps 1 and 2 on the new set of data points, which is the previous set of data points minus the two clusters formed in step 2. Else, go to step 4

4. If there are between $k$ and $2k-1$ points, form a new cluster with these points

5. If there are less than $k$ points then, compute the centroids for all the clusters created so far and find the cluster whose centroid is closest to the centroid of the remaining $k$ points and add the $k$ points to that cluster

However, applying MDAV directly to high dimensional sequential data such as physical activity data results in very high computational cost. The proposed clustering method MC is based on MDAV. However, MC handles high dimensionality of the physical activity data by aggregating the sequences to different time intervals. This improves the efficiency of the anonymization process. In MC (shown in Algorithm 1) all the sequences are assigned to one cluster at the root level (line 1). The sequences are then aggregated to certain time intervals (line 6), for example daily intervals, and then clustered using MDAV (line 7). In the next level, the sequences are drilled down to smaller time intervals (for example, hourly intervals) and clustered using MDAV. These steps are repeated until each cluster at the leaf level has at least $k$ sequences in it. $k$ represents the desired level of anonymity. For clustering, weighted euclidean distance is used for distance computation, with different weights for each activity.

Complexity of MDAV is $O(n^2 m)$ (Domingo-Ferrer et al., 2006b), where $n$ is the number of data points and $m$ is the number of dimensions. The complexity of MC at the root level is $O(n^2 m_L)$ since all the sequences are contained in one cluster. $L$ represents the root level and $m_L$ represents the number of dimensions at the root level ($m_L \ll m$). At the $i^{th}$ level, if $n_i$ represents the size of each cluster, and $m_i$ represents the number of dimensions, then, the complexity is $O(\frac{n}{n_i} n_i^2 m_i)$, which is $O(n n_i m_i)$. There are $L$ levels, so the complexity is given by $O(n \sum_{i=0}^{L} n_i m_i)$, where $L$ represents the root level and 0 represents the leaf level ($n_L = n$).

MC is far more efficient than MDAV for two reasons. At the higher levels, data is aggregated to larger time intervals ($m_i << m$). At lower levels, data is aggregated ($m_i < m$), and additionally, the size of the clusters decreases ($n_i << n$). The quality of the clusters is preserved because, sequences that are similar to one another on a higher level are assigned to the same cluster early on in the clustering process.

The illustration shown in Figure 1 has eight sequences corresponding to physical activity data of eight students at 15-minute interval. For instance, let the k anonymity requirement be 2. At the root level, all the students are assigned to a single cluster. Sequences are aggregated to daily intervals and clustered using MDAV. The clustering results in U1, U2, U3 and U4 being assigned to one cluster, and U5, U6, U7 and U8 to another. In the next level, the sequences are aggregated to hourly intervals and further clustered. The clustering stops since each cluster has at least 2 sequences in it.

---

**Algorithm 1:** Multi-level Clustering

---

**Data:** Set of all sequences $S$, number of levels $l$, aggregation at each level $\{a_1, a_2.., a_l\}$, partition size at each level $\{s_1, s_2.., s_l\}$, required anonymity $k$

**Result:** Set of clusters of sequences $C$

1   $C \leftarrow \{c_1\}$;                /* $c_1$ represents cluster of all sequences */
2   $t \leftarrow 1$;                    /* $t$ is a loop variable */
3   $R \leftarrow NULL$;               /* $R$ stores generated clusters at a level */
4   **while** $t <= l$ **do**
5    **for** $c$ in $C$ **do**
6      Aggregate all the sequences in $c$ to level $a_t$;
7      Using MDAV and $distEuc(X, Y)$ as the distance measure,
       cluster the sequences into partitions of size $s_t$;
                     /* Let $x$ be the number of clusters generated */
8      $R \leftarrow R \cup \{c_{t1}, c_{t2}, ... c_{tx}\}$ ;
9      $C \leftarrow C - \{c\}$;
10    $C \leftarrow R$;
11    $R \leftarrow NULL$;
12    $t \leftarrow t + 1$;
13 **return** $C$;

---

1   $distEuc(X, Y)$

/* X and Y represent two multivariate sequences with n dimensions and m variables */
/* $w_i$ represents weight for each variable (running, walking, and, stationary) */

$$D \leftarrow \sum_{i=1}^{m} w_i \cdot \sqrt{\sum_{j=1}^{n}(x_{ij} - y_{ij})^2}$$      /* m = 3 (running, walking, and stationary) */

2 **return** $D$;

---

*Figure 1. Illustration of Multi-level Clustering*

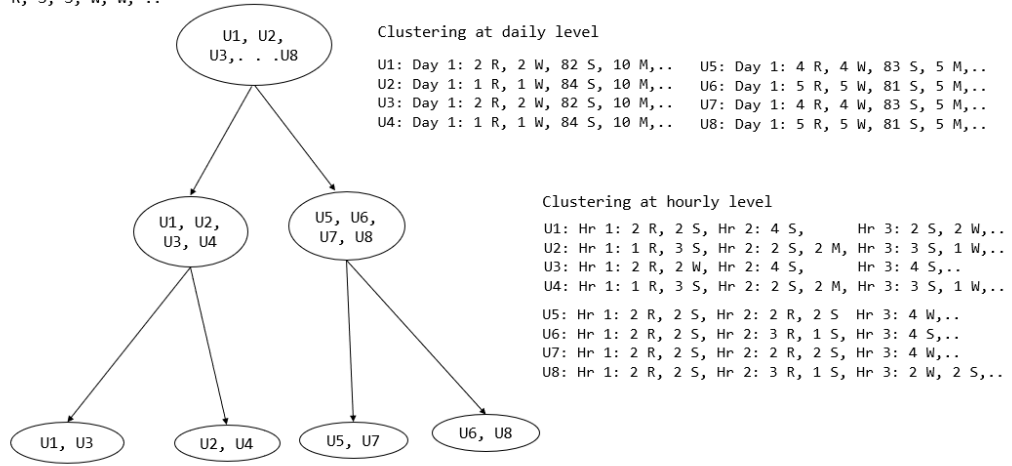

## Multi-level Clustering Based K-Anonymity

*K*-anonymity is a privacy model which ensures that the quasi-identifiers identifying each person cannot be distinguished from at least $k-1$ other individuals (Sweeney, 2002b). In microaggregation, each cluster is replaced by its centroid to achieve k-anonymity. Similar step is performed in MCKA. However, instead of replacing the cluster with the centroid, the centroid is used to simulate sequences. The number of sequences simulated is equal to the size of the cluster. Simulation is done using probabilistic sampling. In Figure 2, the centroid for the cluster of two sequences is computed. Furthermore, using the centroid for probabilistic sampling, two sequences are generated.

*Figure 2. Illustration of Multi-level Clustering Based K-Anonymity*

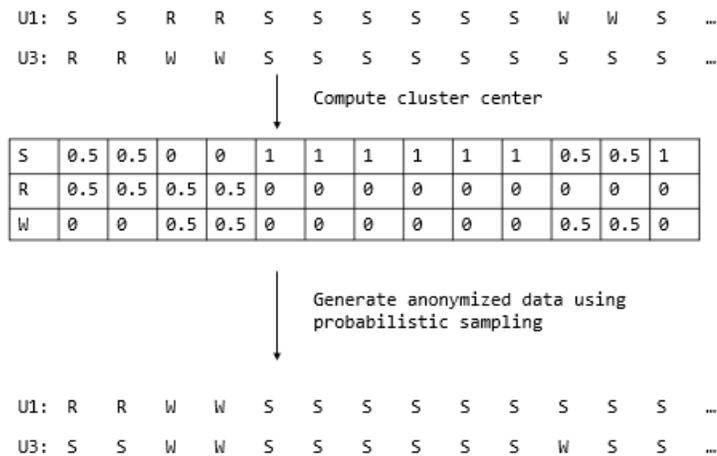

## Multi-level Clustering Based Differential Privacy

Differential privacy is a privacy model that ensures that the presence or absence of an individual's data in the dataset does not make an outcome more or less likely (Dwork, 2008). Differential privacy is stronger than k-anonymity because it can hide the impact of a single person even if the adversaries know the rest of the data set. A standard technique used to achieve differential privacy is the Laplace Perturbation Algorithm (LPA) (Dwork et al., 2006), which works by adding noise to the query results. Let $Q$ be a sequence of queries $Q = \{Q_1, Q_2 \ldots\}$ and the answers to the queries be represented by $Q(I) = \{Q_1(I), Q_2(I) \ldots\}$ where $I$ represents a dataset. In our case, $Q_i$ is a query such as average time of an activity (stationary, walking, running), at an interval *t*, for a given cluster. If a set of queries request for the average time of each activity, at each time interval over the entire time period, then the answers to these queries together represent the centroid of the cluster. Each answer is perturbed by the addition of Laplace noise *Lap(λ)*, therefore, with the increase in the number of queries, noise added also increases.

MCDP uses Fourier Perturbation Algorithm (FPA) (Rastogi & Nath, 2010), which greatly reduces noise to be added. FPA is based on compressing the answers of the query sequence by using orthonormal transformation such as Discrete Fourier Transform (DFT) (Ahmed, Natarajan, & Rao, 1974). The compressed sequence has fewer dimensions as compared to the original sequence, therefore, the overall amount of noise added is reduced. Steps involved in FPA are shown below (Rastogi & Nath, 2010).

1. Compute $C$, DFT of $Q(I)$
2. Consider first *l* co-efficients of $C$, $C^l$
3. Add noise, $\overline{C^l} = C^l + Lap(\lambda)$
4. Pad $\overline{C^l}$ with *0s* ($\overline{C^l_{pad}}$), so the length is same as $C$
5. Compute Inverse DFT (IDFT) of $\overline{C^l_{pad}}$

Noise added is given by $\lambda = \frac{\sqrt{l} \Delta_2(Q)}{\epsilon}$, where $\Delta_2(Q)$ is the $L_2$ sensitivity of $Q$. Sensitivity is the maximum amount the query answers can change when data changes by a row i.e. maximum distance between vectors $Q(I)$ and $Q(I')$ given by $\max|Q(I) - Q(I')|_p$ where $p = \{1,2\}$. $p = 2$ represents $L_2$ distance metric i.e. euclidean distance and the corresponding sensitivity is called $L_2$ sensitivity.

In the proposed approach, to apply FPA, DFT of the centroid is computed, and $Lap(\lambda)$ is added to the first 14 ($l$=14) co-efficients. In this case, $\Delta_2(Q) = \sqrt{m} \cdot \frac{a}{s}$ and $\lambda = \frac{\sqrt{l}\Delta_2(Q)}{\epsilon}$, where $a$ is the size of each interval ($a = 1$), $s$ is size of the cluster, $m$ is the maximal difference (in terms of intervals) between two sequences. Assuming that a sequence can differ at most by 24 hrs per day, and since we have two weeks of data, m =24*60*14=20160. ε is privacy budget and it is set to 1. After noise addition, the sequence is padded with 0s and IDFT is computed. This sequence can then be used to answer the queries. Similar to MCKA, based on the probability of each activity in each interval of the perturbed centroid, as many sequences as the size of cluster is generated and these sequences satisfy ε-differential privacy. Figure 3 shows the centroid for the two sequences U1 and U3, DFT of the centroid, addition of Laplace noise, reconstruction of the centroid and sequences simulated using the noisy centroid.

*Figure 3. Illustration of Multi-level Clustering Based Differential Privacy*

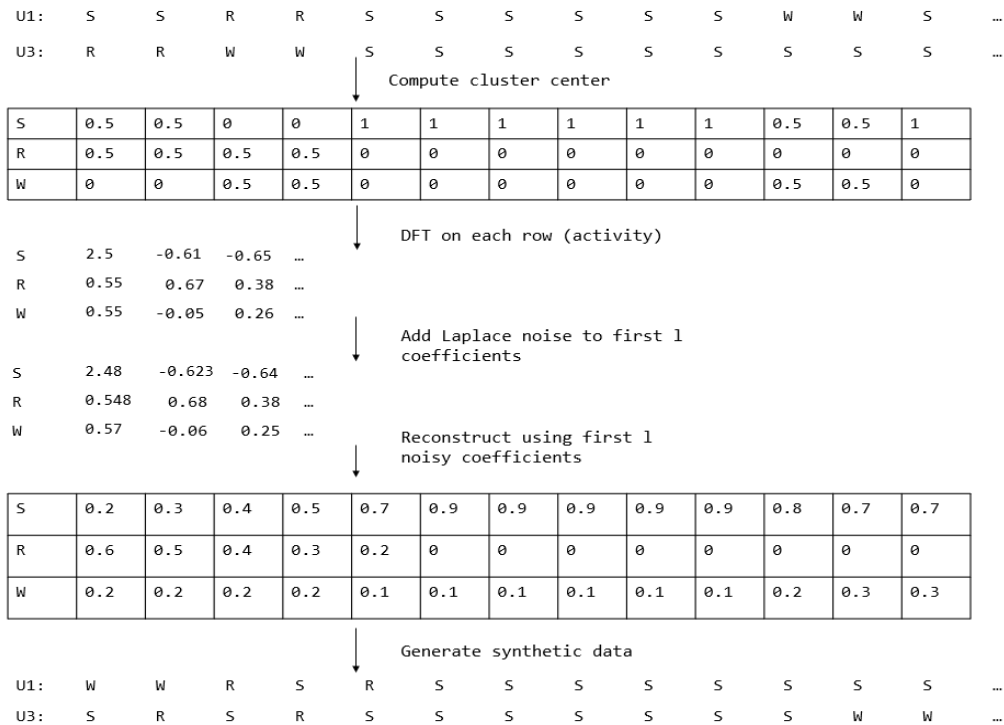

# EXPERIMENTS AND RESULTS

This section starts by describing the experimental set up. It is followed by the experiments for determining the optimal parameters for proposed methods (MC, MCKA and MCDP). Next, the proposed approach is compared with an existing approach as baseline for efficiency and data utility. This is followed by the description of the scaling capabilities of the proposed approach.

## Experimental Setup

### System Requirements

The experiments were conducted on a computer with 32 GB Random Access Memory (RAM) and 3.3 Gigahertz (GHz) processor running the Windows 10 operating system. R was used for programming and all the experiments were single thread implementations.

### Data

For the experiments, synthetic data generated from student life data set was used (refer to the Data subsection in Methods section for a detailed explanation). The synthetic data used in the experiments consists of 9800 students' data (1.9 GB). This data contains activity information for every one-minute interval, for two weeks, for each student. The dimensions of the data set are 9800 rows and 20160 columns.

### Algorithms

The algorithms compared in the experiments are as follows:

1. MCKA: this is the proposed approach which first clusters activity sequences using Multi-level Clustering (MC). This step is followed by applying k-anonymity model by computing centroid for each cluster and using the centroid to generate activity sequences.

2. MCDP: this is the proposed approach which clusters activity sequences using MC followed by implementation of differential privacy. Fourier Perturbation Algorithm (FPA) is used to implement differential privacy.

3. MDAV-KA: this is a baseline method that uses MDAV (Domingo-Ferrer et al., 2006b; Solanas, Martinez-Balleste, & Domingo-Ferrer, 2006) to cluster activity sequences and implements k-anonymity model afterwards.

4. MDAV-DP: this is also a baseline method that uses MDAV to cluster activity sequences and implements differential privacy model afterwards.

The MDAV method ran out of memory over the synthetic data set at minute-level, so the data was aggregated to daily level before running MDAV-KA and MDAV-DP.

### Metrics

Execution time was used to compare the efficiency of the proposed approach with the baseline method. Each method was run five times and the average execution time is reported. The time complexity of the proposed and the baseline approach are presented in the Methods section.

In terms of privacy metrics, $k=5$ was used for MCKA and MDAV-KA where $k$ measures the degree of privacy protection for k-anonymity model. For the two methods implementing differential privacy model (MCDP and MDAV-DP), privacy budget was set to $\varepsilon=1$.

Utility of anonymized data is also important for an anonymization method because anonymized data needs to be useful. Utility was measured by the following metrics:

1. Relative difference

   Relative difference between un-anonymized data and anonymized data was used as the metric for evaluating the effectiveness of the proposed approach. Relative difference (Törnqvist, Vartia, & Vartia, 1985) is computed as

   $$d(x, y) = \frac{|x - y|}{\max(x, y)}$$

   x and y represent un-anonymized and anonymized activity sequence, respectively. Entropy-based metrics (Bayardo & Agrawal, 2005; Sweeney, 2002a; Sweeney, 2001) for evaluating utility were not used since the proposed approach did not involve generalization and suppression-based anonymization. Relative error was not used because values of x or y are often zero.

2. t-test and Cohen's d (effect size)

   Relative difference was computed for both proposed approach and the baseline method. These relative differences were then compared for any significant differences between their means using t-test. A t-test represents whether the difference between different methods is statistically significant. However, it does not tell anything about the size of the effect. Therefore, in addition to the p-values, effect sizes are also reported. Effect size was calculated using Cohen's d value (Cohen, 2013).

3. Correlations

   Pearson's correlation values between activity-flourishing scale, and activity-CGPA were computed for both un-anonymized and anonymized data. This shows whether anonymization preserves correlations in the data.

## Parameter Setting

In this section, the experiments for empirically determining the optimal parameters for MC, MCKA and MCDP are presented. These parameters include 1) fan-out at intermediate level, 2) number of records in a leaf node, 3) number of levels, 4) aggregation of time intervals at each level, and 5) weights for Euclidean distance. The changes in the relative difference are used to determine the optimal values of the parameters. It should be noted that they can vary based on the domain and requirements.

### *Optimal Fan-out at Intermediate Level*

Let $k$ be the number of sequences in a leaf node (these nodes represent the final clusters). The number of records in an intermediate node is $k*p^l$, where $l$ is the level (leaf node has $l=0$, penultimate node (one level above the leaf level) has $l=1$ and so on) and $p$ is the fan-out in the non-leaf level. Experiments were conducted for different values of $p$ (2, 10, 50 *and* 250) with $k=5$. Two levels were used. At the root level, sequences were aggregated to the entire period spanned by the data and at the penultimate level, sequences were aggregated to daily intervals. The relative differences between the un-anonymized and anonymized sequences for *running* activity using MCKA are shown in Figure 4a. Figure 4b shows the time taken for clustering, for different values of *p*.

Relative difference decreases slightly as p increases. This could be because larger value of p (fan-out) leads to larger clusters at intermediate level, which may lead to better clusters in the subsequent levels. Time initially decreases with increase in *p* value. For a very small value of *p* such as *p*=2, a large number of clusters need to be created, which is time consuming. As *p* increases, number of clusters to be computed decreases but after $p = 50$, as the clusters get bigger, computing clusters for consecutive level results in increased time. From the graphs, the optimal value of $p$ is 50 because, at this value of $p$, time taken for clustering is minimum and the relative difference is very close to the minimum value.

*Figure 4: Graphs for determining optimal fan-out at intermediate level*

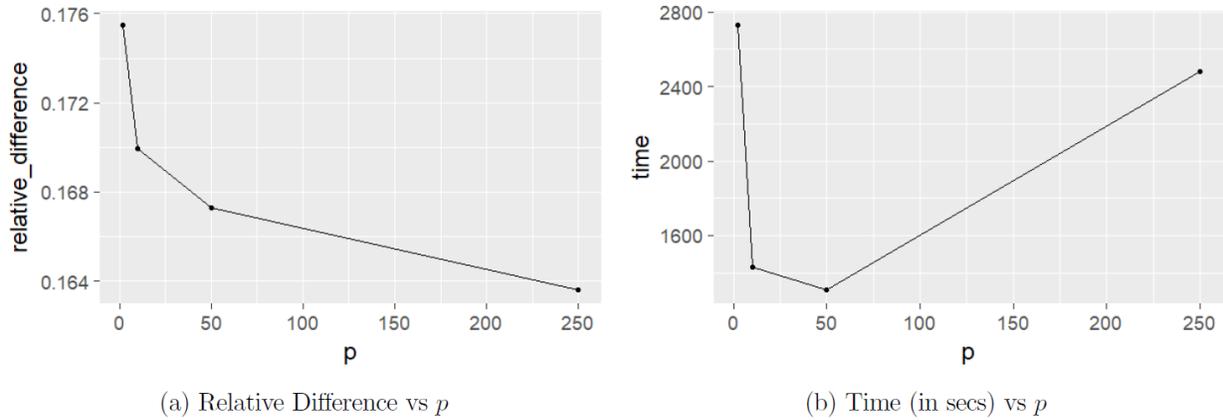

(a) Relative Difference vs $p$

(b) Time (in secs) vs $p$

## Optimal Number of Records in a Leaf Node (k)

To determine the optimal value of $k$, experiments were conducted for different values of $k$ (5, 10, 50, 100), with two levels of clustering. Number of records in an intermediate level node is the minimum of $k*50$ ($p = 50$ is the optimal setting for p) and $N/2$ (there should be at least two intermediate level nodes). Relative differences for *running* activity using MCKA and MCDP are reported.

For MCKA (Figure 5a), the relative difference increases with the increase in $k$ because, homogeneity of the clusters decreases with increase in $k$. For MCDP (Figure 5b), the relative difference decreases with the increase in $k$ initially, because noise-to-be-added decreases with the increase in the size of cluster. After $k = 50$, relative difference slightly increases because of the decreased homogeneity of the clusters. The optimal value of $k$ for MCKA is 5 and optimal value of $k$ for MCDP is 50.

*Figure 5: Graphs for determining optimal number of records in a leaf node*

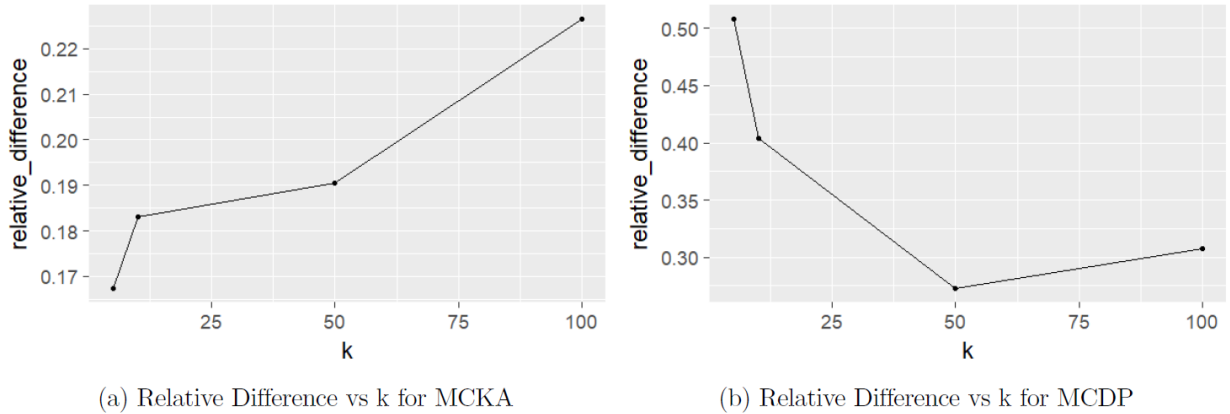

(a) Relative Difference vs k for MCKA

(b) Relative Difference vs k for MCDP

*Optimal Number of Levels in MC*

Experiments were conducted with two and three levels in MC. At the root level, the sequences were aggregated to the entire time duration and at all the other levels, they were aggregated to daily intervals. $k=5$ for a leaf node and number of records at an intermediate level node is the minimum of $k*50^l$ and $N/2$. Relative difference (for MCKA, for *running*) was very similar for two and three levels. However, time taken for clustering increased with the increase in the number of levels (1,280 seconds for two levels and 1,951 seconds for three levels). The optimal number of levels is 2.

*Optimal Aggregation of Time Intervals at Each Level*

To determine optimal aggregation, three experiments with two levels were conducted. At the root level, sequences were aggregated to the entire time duration and at the penultimate level they were aggregated to daily, hourly and 15-minute intervals in each of the three experiments, respectively. Relative difference does not change much for different aggregations (Figure 6a). However, the time taken for clustering decreases drastically with the increase in the aggregation (17651 seconds for 15-min intervals and 1280

*Figure 6: Graphs for determining optimal aggregation*

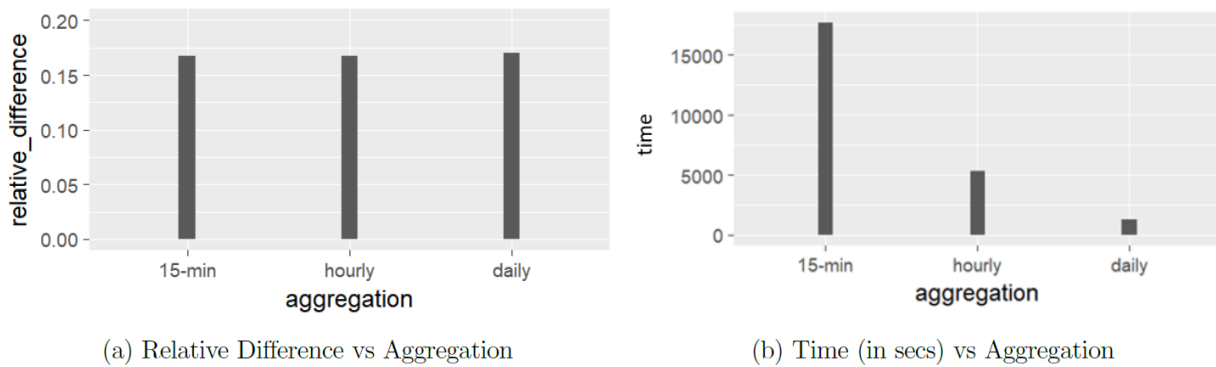

(a) Relative Difference vs Aggregation

(b) Time (in secs) vs Aggregation

seconds for daily intervals) (Figure 6b). Therefore, optimal aggregation is entire-time-duration at the root level and daily interval at the penultimate level.

*Weights for Euclidean Distance*

Two experiments were conducted, one with equal weights for stationary, walking, running and missing, and another with weights in the ratio 1, 50, 150, 50. Running, walking and missing were given higher weights as compared to stationary because these states were infrequent in the data set. Relative difference (for running using MCKA) did not change much for equal weights and ratio-based weights. Therefore, equal weights were used for stationary, walking, running, and missing respectively.

**Evaluation**

*Efficiency*

The execution time of various anonymization algorithms is dominated by the time for clustering. So, using the optimal values for the parameters, experiments were conducted to compare the efficiency of MC (Multi-level clustering, which is used in MCKA and MCDP) and MDAV (clustering method used in MDAV-KA and MDAV-DP). MDAV ran out of memory when applied directly without aggregation (Table 4, column

4). Therefore, sequences were aggregated to daily level for MDAV (consistent with the aggregation in MC). Time taken for clustering (averaged over five runs) are shown in Table 4. The proposed approach took 21 minutes to complete, whereas the baseline method took more than two hours. In all subsequent experiments, MDAV-KA and MDAV-DP have their data aggregated to daily level.

*Table 4. Time for clustering*

|  | MC (k=5) | MDAV (k=5) (with aggregation) | MDAV (k=5) (No aggregation) |
|---|---|---|---|
| **Time for clustering** | 21 mins | 2.6 hrs | >12.6 hrs (memory issues) |

## Data Utility

Total duration of each activity on each day was computed for each un-anonymized activity sequence and its corresponding anonymized activity sequence. The relative difference between them was used to evaluate data utility. Lower value of relative difference means that they are close to each other. Hence, the lower the value, the higher the data utility of the anonymized data.

Table 5 reports the average relative difference for different activities. For example, using MCKA, anonymized data had on average 0.08 relative difference on daily duration of being stationary from un-anonymized data. Running and walking had higher relative difference as compared to stationary because they were infrequent (less than 3% of total time).

*Table 5. Average Relative Difference Per Day (Daily) (S stands for stationary, W stands for walking, R stands for running and M stands for Missing)*

|  | MCKA k=5 | MDAV-KA k=5 | MCDP k=50 | MDAV-DP k=50 |
|---|---|---|---|---|
| **Daily (S)** | 0.08 | 0.08 | 0.14 | 0.14 |
| **Daily (W)** | 0.23 | 0.22 | 0.35 | 0.35 |
| **Daily (R)** | 0.17 | 0.16 | 0.27 | 0.27 |
| **Daily (M)** | 0.22 | 0.21 | 0.42 | 0.42 |

MCKA and MCDP had average relative difference comparable with that of MDAV-KA and MDAV-DP, respectively. Nevertheless, to check whether the difference is statistically significant between the proposed approach and the baseline, t-test was conducted. The t-test was conducted between the relative difference values from MCKA and MDAV-KA. It was also conducted between the relative difference values from MCDP and MDAV-DP. Cohen's d values were also computed to compare the effective size of the difference.

Table 6 shows the p-values and Cohen's d values. Under k-anonymity model (MCKA and MDAV-KA), p-values are significant. This means that, difference between MCKA and MDAV-KA is statistically significant. However, since Cohen's d is less than 0.1 the difference is considered small (Cohen, 2013). Under differential privacy model, the difference between MCDP and MDAV-DP is mostly not statistically significant except for missing values. The results mean that our proposed methods (MCKA and MCDP) lead to similar or slightly worse data utility than the existing methods (MDAV-KA and MDAV-DP), but they are far more efficient than existing methods (see Table 4 where our methods achieved over 5 times speedup even if MDAV aggregates data). Given that physical activity data has very high dimensions, our methods achieve better performance-utility tradeoff than existing methods.

*Table 6. Result of the t-test and Cohen's d between relative difference values from the proposed and the baseline approach*

|  | MCKA & MDAV-KA | | MCDP & MDAV-DP-KA | |
|---|---|---|---|---|
|  | p-value | Cohen's d | p-value | Cohen's d |
| **Daily (S)** | 1.535e-06 | 0.07 | 0.35 | 0.01 |
| **Daily (W)** | 5.309e-07 | 0.07 | 0.16 | 0.02 |
| **Daily (R)** | 0.00014 | 0.05 | 0.38 | 0.01 |
| **Daily (M)** | 1.396e-08 | 0.08 | 0.0001 | 0.05 |

In addition, as an indicator of preserving utility, experiments were conducted to compare activity-flourishing scale (self-perceived success) correlation and activity-CGPA correlation. Pearson's correlation co-efficient for the un-anonymized and anonymized data are shown in Table 7. This shows that the proposed approach preserves both direction and magnitude of the correlations after anonymization.

*Table 7. Pearson's correlation co-efficient for Activity-CGPA*

|  | Activity-Flourishing Scale | | Activity-CGPA | |
|---|---|---|---|---|
|  | Correlation ($r$) | p-value | Correlation ($r$) | p-value |
| **Un-anonymized data** | 0.146 | <2.20E-16 | -0.289 | <2.20E-16 |
| **MCKA (k=5)** | 0.146 | <2.20E-16 | -0.290 | <2.20E-16 |
| **MCDP (k=50)** | 0.129 | <2.20E-16 | -0.293 | <2.20E-16 |

## Scalability

To demonstrate the scaling capabilities, MC was conducted for different number of students (2450, 4900, 7350, 9800) and different duration of the data (one-day, one-week, two-weeks, one-month). The results are shown in Figure 7a and Figure 7b. It can be seen that the algorithm scales linearly. This is because MC spends most time on clustering in lower levels due to increased dimensionality and at lower level $n_i$ (size of cluster) is very small and the cost of clustering is almost linear to $n$ (number of rows) and $m_i$ (number of dimensions).

*Figure 7: Scalability of MC*

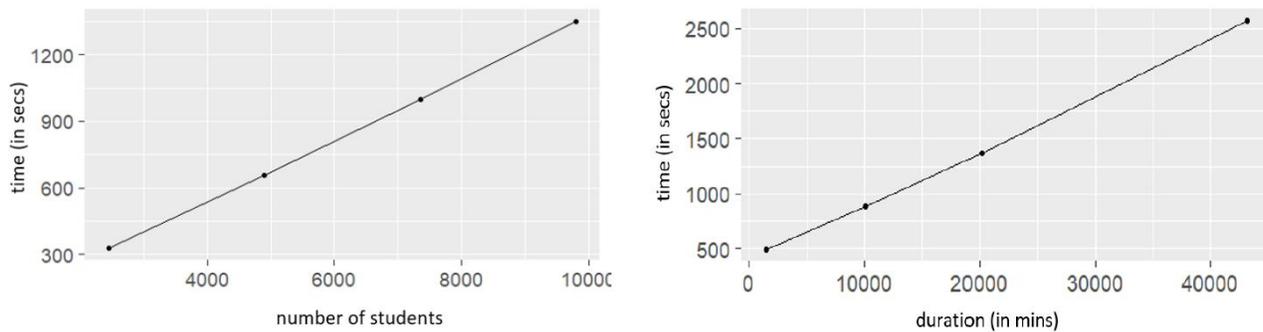

(a) Scaling with Number of Students

(b) Scaling with Duration of the Data

## DISCUSSION

The usefulness of any anonymization technique relies greatly on its capability of preserving data utility. This is important because, if neglected, research studies performed using anonymized data might result in inaccurate and unreliable results. This would in turn defeat the purpose of data publishing. In this study, relative difference between un-anonymized and anonymized data is used to demonstrate the utility preserving capability of the proposed approach. In addition, correlation analysis conducted demonstrates the utility preserving capabilities of the approach. Table 5 shows that MCKA and MCDP have relative difference comparable with MDAV-KA and MDAV-DP. Correlation between activity and flourishing scale for un-anonymized data shows significant positive correlation, and the correlation between activity and CGPA for un-anonymized data shows significant negative correlation (Table 7). The proposed approach preserves both direction and magnitude of the correlations after anonymization. The results show that the proposed approach keeps the utility of the data set intact.

With the increase in the use of wearable sensors, the magnitude of the data collected is increasing dramatically (Banaee, Ahmed, & Loutfi, 2013). This requires the anonymization algorithms to be highly efficient. There are existing approaches that deal with large-scale data anonymization (Zhang et al., 2014). However, they do not take into consideration sequential data sets. Table 4 shows that MC reduces computation time from hours to minutes when compared to the conventional clustering technique such as MDAV (with the proposed dimensionality reduction). Without the dimensionality reduction, MDAV runs into memory issues.

Theoretically we show that our approach has complexity significantly lower than MDAV due to two reasons: 1) at higher level our method aggregates data to very low dimensions; 2) at lower level our method only needs to further divide small clusters rather than the whole data set. In practice, our study shows that it is possible to achieve drastic speed up with the cost of slight drop in data utility when anonymizing large scale physical activity data. These findings agree with existing studies on large scale data anonymization (Zhang, Liu, Nepal, Yang, Dou, & Chen, 2014; Zhang, Dou, Pei, Nepal, Yang, Liu, & Chen, 2014), where they also achieve significant speed up with little loss of data utility. The main difference though is that their approach considers cross-sectional data while our approach focuses on sequential physical activity data. In addition, their approach uses Map-Reduce, which can be an interesting future research direction for our approach as at lower level clusters can be divided into smaller clusters in parallel.

The proposed approach protects privacy while preserving utility, and it speeds up the process drastically as compared to the conventional methods. Therefore, the authors recommend the use of the proposed approach for publishing physical activity data. When published data is used for large number of data points, MCDP can be used because, in MCDP, as the number of data points increases the noise added decreases. MCKA is suitable for both small and large data sets. It preserves utility better than MCDP but provides weaker privacy protection because *k*-anonymity is a weaker privacy model than differential privacy. Both MCKA and MCDP involve the same steps for clustering and only differ in the final step, which is the centroid generation. Therefore, data curators can reuse most of the code for MCKA to compute MCDP and vice-versa.

Like any study, there are certain limitations: 1) Due to unavailability of demographic information in the data set, the study lacks the analysis at the sub-population levels. 2) Synthetic data is used for demonstrating the approach and the original data set contains information for only three types of activities. However, in real-life there are more day-to-day activities such as cycling, climbing and sleeping, that could be considered. 3) Parameter values proposed are suitable for the data set used and might vary depending on the domain and requirements. The last two limitations can be overcome by applying the technique to different data sets. The authors leave this challenge as a part of the future work. 4) Dataset used in this study focuses on student population, however, health related issues are generally associated with elderly population (over the age of 65) (Yanco & Haigh, 2002). Nevertheless, it is worth noting that mental health problems associated with students itself is a highly significant problem in the United States, in terms of severity and number (Hunt & Eisenberg, 2010). The proposed approach is generic enough to be extended to other similar data sets, for example, those related to the activities of the elderly.

In addition, NIH (National Institutes of Health) data sharing policy (National Institutes of Health, 2003) supports and endorses data publication in order to enable free flow of information. However, organizations might refrain from publishing data in order to avoid HIPAA (Health Insurance Portability and Accountability Act of 1996) non-compliance penalties. The proposed approach can enable organizations to follow the encouragements stated in the NIH data sharing policy.

## CONCLUSION

This paper presents a multi-level clustering approach that addresses the issue of high dimensionality in anonymizing sequential data sets such as physical activity data. K-anonymity and differential privacy are then applied to the resulting clusters in order to ensure privacy protection. Compared with the conventional methods, MC based anonymization improves time complexity by reducing the clustering time drastically (from hours to minutes) as compared to MDAV. Both MCKA and MCDP show results (in terms of relative difference) comparable with privacy models applied to conventional clustering technique such as MDAV. The proposed approach also preserves correlation between activity and flourishing scale, and activity and CGPA. In addition, as a part of the future work, the authors intend to apply the same approach to different health related sequential data sets to verify the results and make potentially generalizable recommendations of optimal parameters for the approach.